\documentclass[twocolumn,aps,showpacs,prl]{revtex4}

\usepackage{graphicx}

\usepackage{epsfig}

\begin{document}

\title{Model of nonadiabatic-to-adiabatic dynamical quantum phase transition
in photoexcited systems }

\author{Jun Chang$^{1,2,3}$, Ilya Eremin$^{4,5}$, Jize Zhao$^{6,7}$ }

\affiliation{$^{1}$College of Physics and Information Technology, Shaanxi Normal
University, Xi'an 710119, China\\
 $^{2}$Max-Planck Institut f\"ur Physik komplexer Systeme, D-01187
Dresden, Germany \\
 $^{3}$ State Key Laboratory of Theoretical Physics, Institute of
Theoretical Physics, CAS, Beijing 100190, China\\
 $^{4}$Institut f\"ur Theoretische Physik III, Ruhr-Universit\"at
Bochum, D-44801 Bochum, Germany \\
 $^{5}$Kazan (Volga Region) Federal University, Kazan 420008, Russian Federation \\
 $^{6}$Institute of Applied Physics and Computational Mathematics,
Beijing 100094, China \\
 $^{7}$Beijing Computational Science Research Center, Beijing 100084,
China }

\pacs{64.60.A-, 63.20.kd, 71.27.+a}
\begin{abstract}
We study the ultrafast dynamic process in photoexcited systems and
find that the Franck-Condon or Landau-Zener tunneling between the
photoexcited state and the ground state is abruptly blocked with increasing
the state coupling from nonadiabatic to adiabatic limits. The blockage
of the tunneling inhibits the photoexcited state from decaying into
the thermalized state and results in an emergence of a metastable
state, which represents an entanglement of electronic states with
different electron-phonon coupling strengths. Applying this model
to the investigation of photoexcited half-doped manganites, we show
that the quantum critical transition is responsible for more than
a three-orders-of-magnitude difference in the ground-state recovery
times following photoirradiation. This model also explains some elusive
experimental results such as photoinduced rearrangement of orbital
order by the structural rather than electronic process and the structural
bottleneck of a one-quarter period of the Jahn-Teller mode. We demonstrate
that in the spin-boson model there exist unexplored regions not covered
in the conventional phase diagram. 
\end{abstract}
\maketitle
\textit{Introduction.$-$} Recent years have witnessed increasing
attention to the photoinduced dynamical phase transition in strongly
correlated systems. Controllable light pulses ultrafastly drive the
colossal changes in optical, electronic and magnetic properties of
materials, in particular, by introducing switches between various
competing phases \cite{Miyano1997Photoinduced,Koshihara90,Nasu,Rini,Caviezel2012Femtosecond,li2013femtosecond,hoshino2013time,chuang2013real,Beaud2009Ultrafast,Forst2011Driving,fausti2011light,Wall2009Ultrafast}.
A photoinduced phase transition is believed to be similar to a thermally
driven one because the photon energy eventually is redistributed among
interacting charge, spin and lattice degrees of freedom, and hence
increases the system temperature \cite{Anisimov1974,Beaurepaire96}.

However, the metastable or ``hidden'' phases distinct from those
found in conventional phase diagrams were reported to be accessed
by photoirradiation rather than thermalization in manganites \cite{Ehrke11,Ichikawa11,Singla2013photoinduced,Tobey2012Evolution},
nickelates \cite{lee2012phase}, organic materials \cite{hoshino2013time,Onda08},
cuprates \cite{fausti2011light} and transition metal complexes \cite{Tayagaki01}.
Furthermore, the study of a temporal phase is obviously out of the
reach of traditional methods, where phase transitions are determined
from the free energy of equilibrium states.

The lighting dynamics in manganites, RE$_{x}$AE$_{x,2-x}$MnO$_{3,4}$
(RE, rare-earth ions, AE: alkaline-earth ions), is especially striking
\cite{Ehrke11,Ichikawa11,Singla2013photoinduced,Tobey2012Evolution,Ogasawara2011}.
For instance, in Nd$_{0.5}$Ca$_{0.5}$MnO$_{3}$, the excited state
ultrafastly returns to the ground state within around 0.6 picoseconds
(ps) after photoirradiation \cite{Matsuzaki09}. Surprisingly, in
Nd$_{0.5}$Sr$_{0.5}$MnO$_{3}$ with the same structure, a photoinduced
long-lived excited state survives for about 3000 ps or 3 nanoseconds
(ns) \cite{Ichikawa11}. Similarly, in La$_{0.5}$Sr$_{1.5}$MnO$_{4}$,
the photoexcited transient state also lasts for ns \cite{Ehrke11}.
The underlying mechanism making the enormous difference in recovery
times is still not clear. Another challenge in the manganites is that
the photoexcitation melts antiferromagnetic order but only partially
reduces the orbital order \cite{Ehrke11}. In addition, a theoretical
understanding of the structural bottleneck and photoinduced rearrangement
of the orbital order by the structural rather than the electronic
process \cite{Singla2013photoinduced} is timely.

In this paper, we focus on the ultrafast quantum phase transition
and the formation of metastable states in the photoexcited spin-boson-like model \cite{Leggett}. With the aid of this model, we probe the
ultrafast local electron dynamics in strongly correlated systems after
low-intensity light irradiation.

\textit{Quantum Model.$-$} In a strongly correlated system, optical
light illumination triggers the excitation of the higher vibrational
levels of phonon modes, and drives a redistribution of anisotropic
$d$ or $f$ orbital occupations. Hence it often leads to geometric
deformation or structural phase transition in the system. The locally
excited state dissipates energy to its surrounding by emission of
phonons or photons \cite{Weiss}. To elucidate this dynamical process,
we introduce a model with electronic states, coupled to a phonon bath.
Due to the strong electron-phonon coupling and the substantial bath
memory effects in a photo-driven system, a Born-Markov master equation
is insufficient to describe the ultrafast electron dynamics. For this
reason, we first map the spin-boson model to an alternative model,
where the electronic states with energies $E_{i}$ are coupled to
a single harmonic mode damped by an Ohmic bath \cite{Garg}. Here,
we assume that the correlations between electrons are taken into account
by the effective renormalization of electronic state energies. Variations
in the coupling strength $\lambda_{i}$ change the equilibrium positions
of different states. The system Hamiltonian is written as, 
\begin{eqnarray}
H_{s}=\sum_{i}E_{i}c_{i}^{\dagger}c_{i}+\sum_{ij}V_{ij}(c_{i}^{\dagger}c_{j}+{\rm h.c.})\nonumber \\
+\sum_{i}\lambda_{i}c_{i}^{\dagger}c_{i}\left(a^{\dagger}+a\right)+\hbar\omega a^{\dagger}a,\label{h0}
\end{eqnarray}
where $c_{i}^{\dagger}c_{i}$ gives the occupation of the state $i$,
$V_{ij}$ is the coupling (hybridization) constant that causes a transition
between states $j$ and $i$. $a^{\dagger}$ is the creation operator
for the vibrational mode with frequency $\omega$. We further define
the energy gap $\Delta_{ij}=\left(E_{i}-\varepsilon_{i}\right)-\left(E_{j}-\varepsilon_{j}\right)$
and electron-phonon self-energy difference $\varepsilon_{ij}=\left(\lambda_{i}-\lambda_{j}\right)^{2}/(\hbar\omega)$
between two states with $\varepsilon_{i}={\lambda_{i}}^{2}/(\hbar\omega)$.
Interestingly, despite $H_{s}$'s formal similarity with the Holstein
model \cite{Holstein}, the indices $i$ and $j$ refer to states
rather than lattice sites.

The Ohmic bath damping is introduced by a dissipative Schr\"odinger
equation, in which a dissipative operator $iD$ is added to the Hamiltonian
to describe the bath induced state transfer, $i\hbar\partial|\psi(t)\rangle/\partial t=\left(H_{0}+iD\right)|\psi(t)\rangle$,
where $H_{0}$ is the Fr\"ohlich transformation of $H_{s}$. This
equation effectively incorporates both the strong electron-phonon
coupling and environment memory effects, which were described previously
\cite{Chang12,Veenendaal10,Chang10}. Here, we focus on the dynamics
of the electronic states with electron number conservation at the low
excitation photon density limit and set the temperature $T=0$, as
we are interested in the quantum phase transitions.

\begin{figure}
\includegraphics[clip,width=0.6\columnwidth]{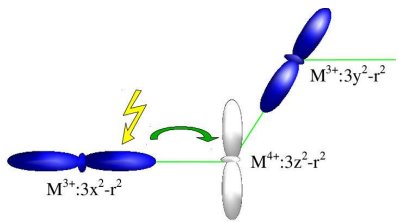} \protect\caption{(color online) Schematic photoinduced charge transfer from $3x^{2}/3y^{2}-r^{2}$
orbitals at Mn$^{3+}$ to $3z^{2}-r^{2}$ and/or $x^{2}-y^{2}$ (not
shown) orbitals at their neighboring Mn$^{4+}$ sites in half-doped
manganites. The (green) line indicates a ferromagnetic zigzag chain.
Before the pump pulse, $3z^{2}-r^{2}$ orbitals at Mn$^{4+}$ sites are
empty. Oxygen orbitals are not shown for clarity. }

\label{Mnchain} 
\end{figure}

Let us specifically consider half-doped manganites as an example.
The insulating ground state of this system is mostly characterized
by the charge-exchange type order with the ferromagnetic (FM) zigzag
chains, which couple antiferromagnetically, whereas Mn$^{4+}$ and
Mn$^{3+}$ alternate on the chain \cite{Jirak85,Sternlieb1996charge,Murakami1998direct}.
An incident optical photon drives an electron transfer from the $3x^{2}/3y^{2}-r^{2}$
orbital at the Mn$^{3+}$ site to the $3z^{2}-r^{2}$ and/or the $x^{2}-y^{2}$
orbital at its neighboring Mn$^{4+}$ site on the FM chain \cite{Singla2013photoinduced,Okimoto99}.
The charge redistribution in the anisotropic $d$ orbitals often occurs
along with strong lattice oscillations. For instance, in Nd$_{0.5}$Ca$_{0.5}$MnO$_{3}$,
the so-called Jahn-Teller mode is dominant to release the displacements
of oxygen atoms around the Mn ions after lighting \cite{Matsuzaki09}.
To apply the above model in the low laser intensity limit (the saturation
density of excitation photons, 0.8 mJ cm$^{-2}$, corresponds to one
photon per 60 Mn ions \cite{Ichikawa11}), we label the initial state
prior to the lighting as state $1$, and the charger transfer state
as $2$. The surrounding sites are regarded as the environment. According
to the optical spectra experiments \cite{Okimoto99}, both the energy
gap $\Delta$ and the electron-phonon self-energy difference $\varepsilon$
between state $1$ and $2$ are estimated to be around several tenths
of eV. This is also in agreement with Jahn-Teller characteristic energy.
We include the Jahn-Teller mode frequency $\hbar\omega$=0.06 eV and
damping time on the picoseconds time scale, $(2\Gamma)^{-1}=0.1$
ps following the reference \cite{Matsuzaki09}. It is worth noting
that the decay time of a particular state is not directly related
to the phonon mode damping $\Gamma$. As for the electronic state
coupling or hybridization parameter $V$, its value ($(pd\sigma)^{2}/2$)
ranges between 0.13 and 0.33 eV in the literature \cite{Dagotto01}
for the $3x^{2}/3y^{2}-r^{2}$ and $3z^{2}-r^{2}$ orbitals. Note,
that $pd\sigma$ is the overlap integral between the $d\sigma$ and $p\sigma$-orbitals.
Here, we also set $\varepsilon=0.4$ eV as the energy unit.

\begin{figure}
\includegraphics[width=1\columnwidth]{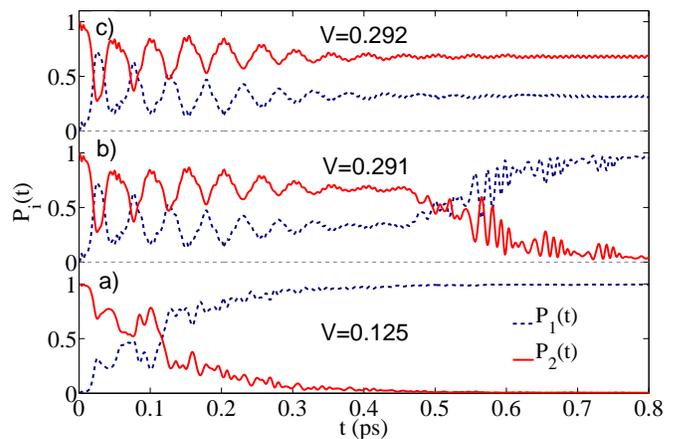} \protect\caption{(color online) The time evolution of photodriven state probabilities
for $V$=0.125 (a), $V$=0.291 (b), and $V=V_{c}$=0.292 (c). When
the electronic coupling $V$ is less than a critical value $V_{c}$,
the excited state $2$ returns to the ground state $1$. While $V\geq V_{c}$,
the relaxed state represents a strong entanglement of both the state
$1$ and the state $2$. The slower oscillation period refers to the
phonon mode of 69 fs, while the faster oscillation period of about
9$-$10 fs is expected from the energy gap $\sqrt{\Delta^{2}+4V^{2}}$.
Here, we set $\Delta=\varepsilon$ and $\varepsilon=0.4$ eV as the
energy unit. }

\label{time} 
\end{figure}

\textit{Quantum Phase Transition.$-$} The first photoexcited state
is set as the starting state, which keeps the same configuration coordinate
as the ground state in the Born approximation. Solving the dissipative
Schr\"odinger equation numerically, the evolution of the photoexcited
state as a function of time clearly reflects the quantum phase transition,
see Fig. \ref{time}. When the hybridization is weak, i.e. $V$=0.125,
far less than the electron-phonon coupling $\lambda=\sqrt{\hbar\omega\varepsilon}$=0.387,
the photoexcited state relaxes to the ground state with quantum efficiency
close to 100\% within 0.5 ps, which is close to the recovery time
0.6 ps, found in Nd$_{0.5}$Ca$_{0.5}$MnO$_{3}$.\cite{Matsuzaki09}
Here, we notice that our results obtained by the quantum method qualitatively
agree with that of the semi-classical Franck-Condon principle \cite{Chang10}.
Increasing $V$ to 0.291, the photodriven state still falls back to
the initial state within 1 ps (see Fig. \ref{time} (b)). However,
once $V$ raises to the critical value $V_{c}$=0.292, the occupation
probability of state $2$ stays finite after around 0.5 ps. An extension
of the time evolution up to 0.1 ns confirms that the long-lived excited
state represents a strongly entangled mixture of states $1$ and $2$,
which does not decay by phonon dissipation. With further increased
hybridization $V>V_{c}$, the metastable state remains robust, which
signals quantum phase transition to the metastable state at $V_{c}$.

\begin{figure}[t]
\includegraphics[width=0.48\columnwidth]{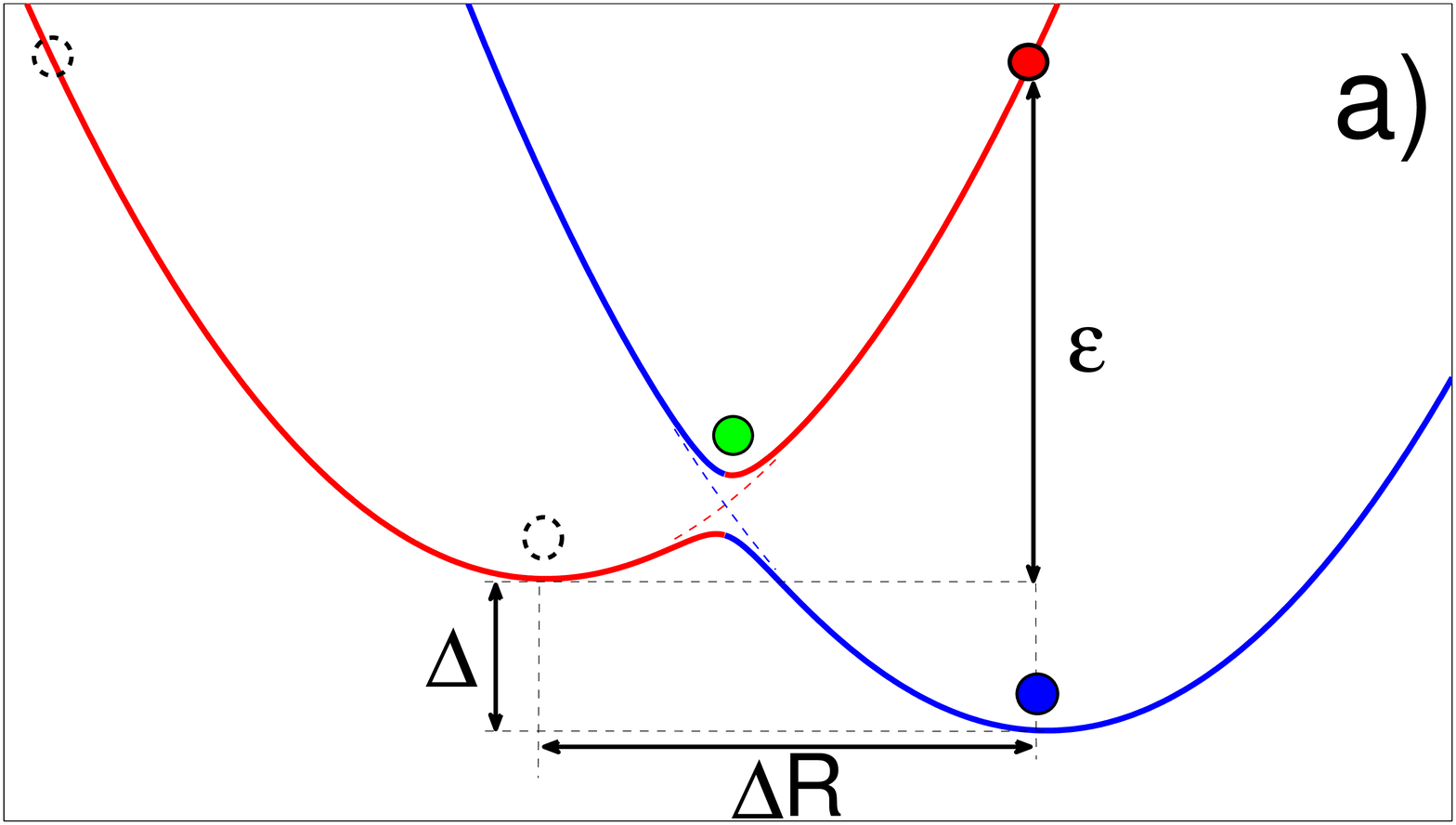} \includegraphics[width=0.48\columnwidth]{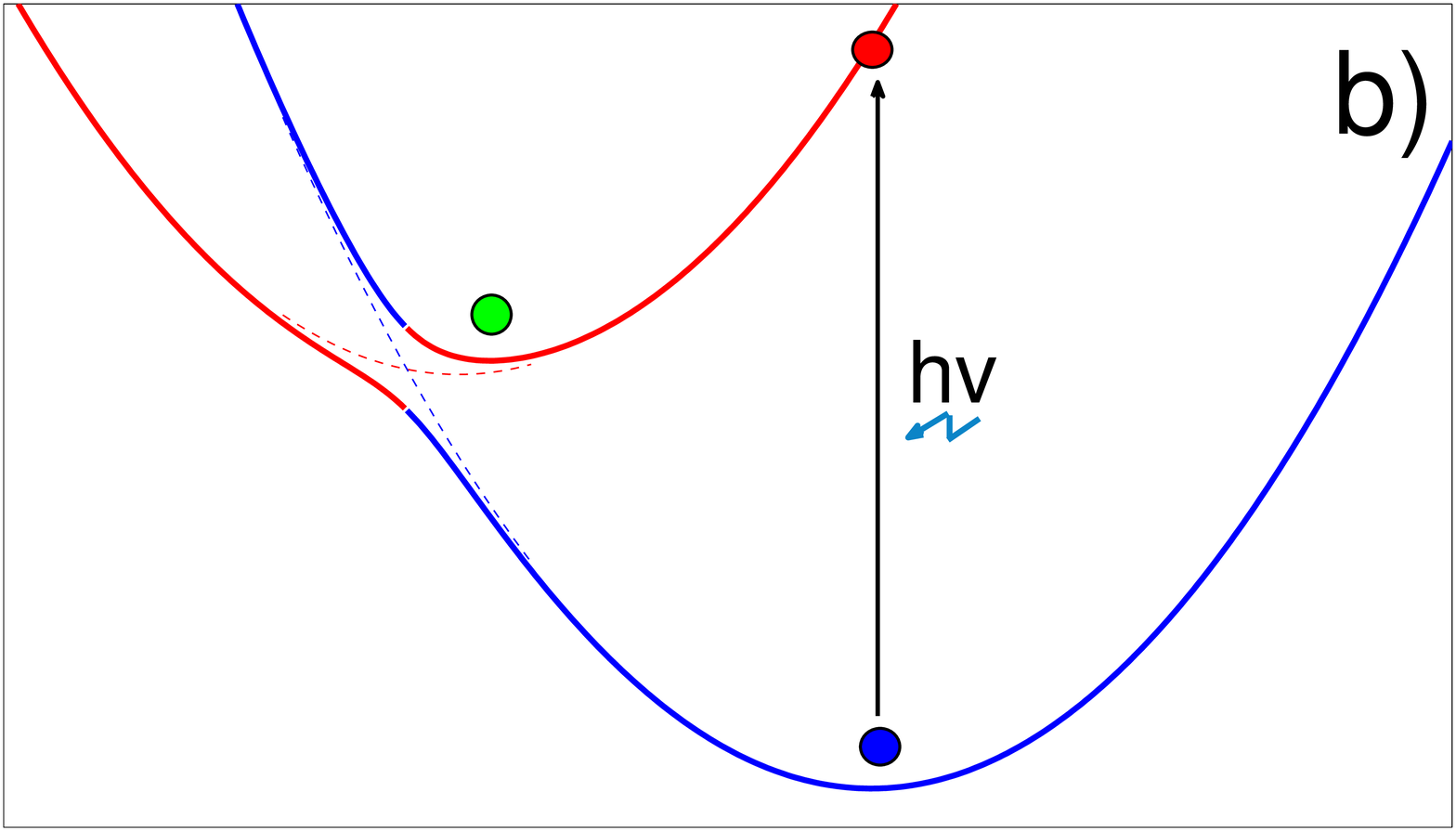}
\protect\caption{(color online) Schematic free energy as a function of the configuration
coordinate for two energy levels, coupled to the harmonic phonon mode.
The parabolic curves are nonadiabatic free energy surfaces in the
weak-coupling limit. The electronic coupling caused splitting at the cross point is neglected. The photoexcited state reaches the ground state
by Franck-Condon or Laudau-Zener tunneling. The solid anticrossing
curves refer to the adiabatic free energy surfaces, and the energy
gap (about $2V$) at the avoided crossing point blocks the tunneling and separates the photoexcited states from the
thermalized states on the lower potential energy curve. Consequently,
the photoinduced states (red dots) relaxe to the bottom of the higher
curves (green dots with energy E$_{c}$). Here, $\Delta$ and $\varepsilon$
are the energy gap and electron-phonon self-energy difference between
the two levels, respectively.}

\label{schematic} 
\end{figure}

\textit{Semi-classical Model.$-$} Here, we present a qualitative
understanding of the metastable state formation, starting from a semi-classical
phenomenological model with two localized electronic states, coupled
to a phonon mode with frequency $\omega$ as shown in Fig.\ref{schematic}.
The crossing curves are parabolic nonadiabatic free-energy surfaces.
The ground-states is labeled as $1$ with the free-energy as a function
of the configuration coordinate $f_{1}=KR^{2}/2+\Delta$, and the
excited state $2$ with $f_{2}=K(R-\Delta R)^{2}/2$. Here, $\Delta$
is the energy gap between two states, and their equilibrium positions
are separated by $\Delta R$. We further define the electron-phonon
self-energy $\varepsilon=K(\Delta R)^{2}/2$ and the corresponding
electron-phonon coupling strength $\lambda=\sqrt{\hbar\omega\varepsilon}$.
The excited state relaxes to the original state via the weak energy
splitting at the free energy surface cross point \cite{yarkony1996diabolical} according to
the Franck-Condon principle or Landau-Zener tunneling in the weak-coupling
limit, i.e. when the strength of the electronic state coupling $V\ll\lambda$.

The solid anti-crossing curves of adiabatic free energy surfaces are
described by $\epsilon_{\pm}=\left(f_{1}+f_{2}\pm\sqrt{\left(f_{1}-f_{2}\right)^{2}+4V^{2}}\right)/2$.
The energy gap near the avoided crossing is about $2V.$ The photoexcited
state relaxes to the bottom of the upper potential energy curve as
shown in Fig. \ref{schematic}, whereas the thermally excited state stays
on the lower energy curve in the adiabatic limit $V\gg\lambda$.

This classical model has been extensively used to explain many experimental
results. For example, Marcus' theory adopts this model to give the
probability of interconversion of donor and acceptor through the region
near the intersection of potential energy surfaces in the non-adiabatic
limit \cite{Marcus}. However, the metastable state often refers to
the thermally induced state in the bottom of the lower adiabatic free
energy surfaces \cite{Marcus,Tisdale11}, represented by the lower
dashed circle in Fig. \ref{schematic}(a). As clearly seen it is distinctly
different from our proposed photoinduced metastable state in the bottom
of the upper anti-crossing curves (green dots). Particularly, there is
no classical metastable state when $\Delta$ is close to $\varepsilon$
(see Fig. \ref{schematic}(b)). Additionally, the nonequilibrium
phase transition is also out of reach of the semi-classic model. We
have shown that a solution of the time dependent Schr\"odinger equation
provides a practical approach to study the dynamical phase transition
process, not restricted to the adiabatic or nonadiabatic limits. 

\begin{figure}
\includegraphics[width=0.9\columnwidth]{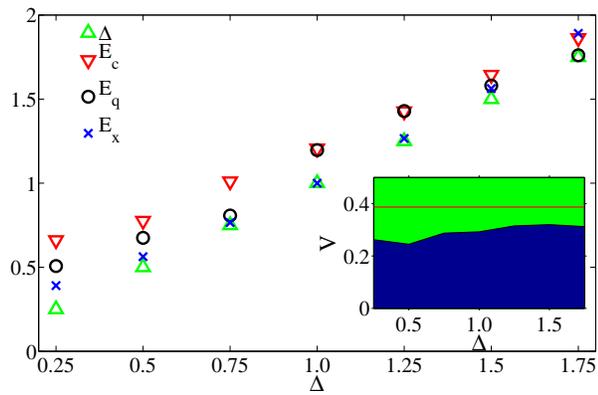} \protect\caption{(color online) The photoinduced metastable state phase diagram. The
energies of the photoinduced quantum metastable state E$_{q}$ are
located below the energies E$_{c}$ at the bottom of the upper adiabatic
curves, shown in Fig. \ref{schematic}, and above the energy gap $\Delta$.
E$_{x}$ is the free energy at the cross point of the nonadiabatic
curves. The inset shows critical strength of the state coupling $V{}_{c}$
for the formation of the metastable state. The (red) straight line
indicates the electron-phonon coupling $\lambda$.}

\label{Ei} 
\end{figure}

\textit{Metastable State.$-$}To verify the robustness of the metastable
state, we vary the energy gap away from the electron-phonon self-energy.
We find that the photoinduced quantum metastable state still exists
and its state energy E$_{q}$ is quite close to the classical free
energy E$_{c}$, shown at the bottom of the upper adiabatic curves
in Fig. \ref{schematic} (green circles). The inset of Fig.\ref{Ei}
shows the critical strength of the state coupling $V_{c}$ for the
formation of the metastable states. However, we find that E$_{q}$
is also close to the energy gap $\Delta$, which is often referred
to as the classical metastable state energy, see Fig. \ref{Ei}. To further
confirm our phenomenological picture for the metastable state, we
prepare some thermalized initial states and study their time evolution
in the dissipative Schr\"odinger equation. For example, turning the
coupling between the states off, we put the starting state to those,
indicated by the dashed circles in Fig.\ref{schematic}, and then turn
the electric coupling on. Regardless of the coupling being larger
or lesser than the critical $V_{c}$, these thermalized states decay
to the ground state within 1 ps without long-lived metastable states,
at least in the parameter space of this paper. This also suggests
the impossibility for thermalization to excite the 'hidden' phases
in experiments.

\textit{Discussion and Conclusion.$-$} Our results for the dynamical
phase transition allow us to clarify some elusive experimental results,
for instance, the three order of magnitude difference in the ground state
recovery times between Nd$_{0.5}$Sr$_{0.5}$MnO$_{3}$ , La$_{0.5}$Sr$_{1.5}$MnO$_{4}$
and Nd$_{0.5}$Ca$_{0.5}$MnO$_{3}$. We believe that the photoexcited
state falls back to the ground state within 1 ps in Nd$_{0.5}$Ca$_{0.5}$MnO$_{3}$
without formation of the metastable state, which could be due to the hybridization
$V$ being less than the critical value $V_{c}$ \cite{Matsuzaki09}.
At the same time, the photoinduced excited state relaxes to the transient
state in the other two compounds \cite{Ehrke11,Ichikawa11}, as the
long-lived excited state does not decay by the phonon, and alternatively,
it relaxes to the initial state by the spontaneous emission of fluorescence
photons with a characteristic relaxation time of 1$\sim$100 ns. As
for the antiferromagnetic order destroyed in layer manganites La$_{0.5}$Sr$_{1.5}$MnO$_{4}$,
we propose here that the charge transfer occupation of the $3z^{2}-r^{2}$
orbitals increases the layer spacing and further reduces the already
weak layer antiferromagnetic coupling, and AF order is consequently
suppressed \cite{Ehrke11}. On the other hand, orbital order is only
partially reduced because a recurrence to the initial state occurs
after the charge transfer from $3x^{2}/3y^{2}-r^{2}$ to $3z^{2}-r^{2}$
within tens of fs as shown in Fig \ref{time}(c). Meanwhile, a drop
time, about a one-quarter period of the Jahn-Teller mode, is obtained
by the error function fit of the charge transfer state probability
evolution with time in the Fig \ref{time}(c). This is in agreement
with the experiment, which observes 18 fs bottleneck for the loss
of the orbital order \cite{Singla2013photoinduced}. The transient
metastable state is then a strong entanglement of the Jahn-Teller
orbital doublet, which is also consistent with the experimental evidence
of photoinduced rearrangements of orbital order by the structural
rather than the electronic process in a recent birefringence experiment
\cite{Singla2013photoinduced}.

To summarize, we have solved the dissipative Schr\"odinger equation
for a two level system in phonon bath to simulate the ultrafast time
evolution of local quantum states in photoexcited strongly correlated
systems. A dynamical quantum phase transition is indicated by the
formation of a metastable state with increasing the state coupling
from nonadiabatic to adiabatic limits. By the semi-classical model,
we show that the emergent transient metastable states could be stabilized
by the blockage of the Laudau-Zener tunneling near the avoided crossing,
distinctly different from the conventional stabilization of transient
charge or orbital order by the on-site or inter-site Coulomb repulsion.
This vibronic mechanism is substantiated by the recent optical birefringence
experiment \cite{Singla2013photoinduced}. The dynamics from nonadiabatic
to adiabatic limit experiences a quantum phase transition rather than
a gradual process. This physical picture also could be extended towards
understanding the photoinduced formation of temporal hidden phases
in other transition metal or rare earth systems \cite{lee2012phase},
organic materials \cite{hoshino2013time,Onda08}, transition metal
complexes \cite{Tayagaki01} and potentially even the systems with Dirac cones. We further expect that
the pressure induced change of electronic states coupling may potentially
drive the quantum phase transition in a single material, e.g. Nd$_{0.5}$Ca$_{0.5}$MnO$_{3}$.
Our work demonstrates that there exist hidden regions in conventional
phase diagrams when correlated systems are driven out of equilibrium.

\textit{Acknowledgments}.$-$We are thankful to Yang Ding, Javier
Fernandez Rodriguez, Michel van Veenendaal, Stefan Kirchner, Peter
Thalmeier, Roderich Moessner and Shaojin Qin for fruitful discussions.
J.C is supported by the Fundamental Research Funds for the Central
Universities, GK201402011, and SNNU's seed funds. The work of I.E.
is supported by the Mercator Research Center Ruhr (MERCUR) and the
Russian Government Program of Competitive Growth of Kazan Federal
University.J.Z. is supported by NSFC 11474029.



\begin{thebibliography}{10}
\bibitem{Miyano1997Photoinduced} K. Miyano, T. Tanaka, Y. Tomioka,
and Y. Tokura, Phys. Rev. Lett. \textbf{78,} 4257 (1997).

\bibitem{Koshihara90} S. Koshihara, Y. Tokura, T. Mitani, G. Saito,
and T. Koda, Phys. Rev. B \textbf{42,} 6853 (1990).

\bibitem{Nasu} K. Nasu, Photoinduced Phase Transitions (World Scientific,
2004).

\bibitem{Rini} M. Rini, R. Tobey, N. Dean, J. Itatani, Y. Tomioka,
Y. Tokura, R. W. Schoenlein, and A. Cavalleri, Nature (London) \textbf{449,}
72 (2007).

\bibitem{Caviezel2012Femtosecond} A. Caviezel, U. Staub, S. L. Johnson,
S. O. Mariager, E. M\"ohr-Vorobeva, G. Ingold, C. J. Milne, M. Garganourakis,
V. Scagnoli, S. W. Huang, et al., Phys. Rev. B \textbf{86,} 174105
(2012).

\bibitem{li2013femtosecond} T. Li, A. Patz, L. Mouchliadis, J. Yan,
T. A. Lograsso, I. E. Perakis, and J. Wang, Nature (London) \textbf{496,}
69 (2013).

\bibitem{hoshino2013time} M. Hoshino, S. Nozawa, T. Sato, A. Tomita,
S.-i. Adachi, and S.-y. Koshihara, RSC Adv. \textbf{3,} 16313 (2013).

\bibitem{chuang2013real} Y. Chuang, W. Lee, Y. Kung, A. Sorini, B.
Moritz, R. Moore, L. Patthey, M. Trigo, D. Lu, P. Kirchmann, et al.,
Phys. Rev. Lett. \textbf{110,} 127404 (2013).

\bibitem{Beaud2009Ultrafast} P. Beaud, S. L. Johnson, E. Vorobeva,
U. Staub, R. A. D. Souza, C. J. Milne, Q. X. Jia, and G. Ingold, Phys.
Rev.Lett. \textbf{103,} 155702 (2009).

\bibitem{Forst2011Driving} M. F\"orst, R. I. Tobey, S. Wall, H.
Bromberger, V. Khanna, A. L. Cavalieri, Y.-D. Chuang, W. S. Lee, R.
Moore, W. F. Schlotter, et al., Phys. Rev. B \textbf{84,} 241104 (2011).

\bibitem{fausti2011light} D. Fausti, R. Tobey, N. Dean, S. Kaiser,
A. Dienst, M. Homann, S. Pyon, T. Takayama, H. Takagi, and A. Cavalleri,
Science \textbf{331,} 189 (2011).

\bibitem{Wall2009Ultrafast} S. Wall, D. Prabhakaran, A. T. Boothroyd,
and A. Cavalleri, Phys. Rev. Lett. \textbf{103,} 097402 (2009).

\bibitem{Anisimov1974} S. I. Anisimov, B. L. Kapeliovich, and T.
L. Perel'man, Zh. Eksp. Teor. Fiz. \textbf{66,} 776 (1974), {[}Sov.
Phys. JETP \textbf{39,} 375 (1974){]}.

\bibitem{Beaurepaire96} E. Beaurepaire, J.-C. Merle, A. Daunois,
and J.-Y. Bigot, Phys. Rev. Lett. \textbf{76,} 4250 (1996).

\bibitem{Ehrke11} H. Ehrke, R. I. Tobey, S. Wall, S. A. Cavill, M.
F\"orst, V. Khanna, T. Garl, N. Stojanovic, D. Prabhakaran, A. T.
Boothroyd, et al., Phys. Rev. Lett. \textbf{106,} 217401 (2011).

\bibitem{Ichikawa11} H. Ichikawa, S. Nozawa, T. Sato, A. Tomita,
K. Ichiyanagi, M. Chollet, L. Guerin, N. Dean, A. Cavalleri, S.-i.
Adachi, et al., Nat. Mater. \textbf{10,} 101 (2011).

\bibitem{Singla2013photoinduced} R. Singla, A. Simoncig, M. Forst,
D. Prabhakaran, A. L. Cavalieri, and A. Cavalleri, Phys. Rev. B \textbf{88,}
075107 (2013).

\bibitem{Tobey2012Evolution} R. I. Tobey, S. Wall, M. F\"orst, H.
Bromberger, V. Khanna, J. J. Turner, W. Schlotter, M. Trigo, O. Krupin,
W. S. Lee, et al., Phys. Rev. B \textbf{86,} 064425 (2012).

\bibitem{lee2012phase} W.-S. Lee,Y. Chuang, R. Moore, Y. Zhu, L.
Patthey, M. Trigo, D. Lu, P. Kirchmann, O. Krupin, M. Yi, et al.,
Nat. Commun. \textbf{3,} 838 (2012).

\bibitem{Onda08} K. Onda, S. Ogihara, K. Yonemitsu, N. Maeshima,
T. Ishikawa, Y. Okimoto, X. Shao, Y. Nakano, H. Yamochi, G. Saito,
et al., Phys. Rev. Lett. \textbf{101,} 067403 (2008).

\bibitem{Tayagaki01} T. Tayagaki and K. Tanaka, Phys. Rev. Lett.
\textbf{86,} 2886 (2001).

\bibitem{Ogasawara2011} T. Ogasawara, T. Kimura, T. Ishikawa, M.
Kuwata- Gonokami, and Y. Tokura, Phys. Rev. B \textbf{63,} 113105
(2001).

\bibitem{Matsuzaki09} H. Matsuzaki, H. Uemura, M. Matsubara, T. Kimura,
Y. Tokura, and H. Okamoto, Phys. Rev. B \textbf{79,} 235131 (2009).

\bibitem{Leggett} A. J. Leggett, S. Chakravarty, A. T. Dorsey, M.
P. A. Fisher, A. Garg, and W. Zwerger, Rev. Mod. Phys. \textbf{59,}
1 (1987).

\bibitem{Weiss} U. Weiss, Quantum Dissipative Systems (World Scientific,
Singapore, 2000).

\bibitem{Garg} A. Garg, J. Onuchic, and V. Ambegaokar, J. Chem. Phys.
\textbf{83,} 4491 (1985).

\bibitem{Holstein} T. Holstein, Ann. Phys. \textbf{8,} 325 (1959).

\bibitem{Chang12} J. Chang, A. J. Fedro, and M. van Veenendaal, Chem.
Phys. \textbf{407,} 65 (2012).

\bibitem{Veenendaal10} M. van Veenendaal, J. Chang, and A. J. Fedro,
Phys. Rev. Lett. \textbf{104,} 067401 (2010).

\bibitem{Chang10} J. Chang, A. J. Fedro, and M. van Veenendaal, Phys.
Rev. B \textbf{82,} 075124 (2010).

\bibitem{Jirak85} Z. Jir\'ak, S. Krupi\'cka, Z. \v Sim\v sa,
M. Dlouh\'a, and S. Vratislav, J. Magn. Magn. Mater. \textbf{53,}
153 (1985), ISSN 0304-8853.

\bibitem{Sternlieb1996charge} B. J. Sternlieb, J. P. Hill, U. C.
Wildgruber, G. M. Luke, B. Nachumi, Y. Moritomo, and Y. Tokura, Phys.
Rev. Lett. \textbf{76,} 2169 (1996).

\bibitem{Murakami1998direct} Y. Murakami, H. Kawada, H. Kawata, M.
Tanaka, T. Arima, Y. Moritomo, and Y. Tokura, Phys. Rev. Lett. \textbf{80,}
1932 (1998).

\bibitem{Okimoto99} Y. Okimoto, Y. Tomioka, Y. Onose, Y. Otsuka,
and Y. Tokura, Phys. Rev. B \textbf{59,} 7401 (1999).

\bibitem{Dagotto01} E. Dagotto, T. Hotta, and A. Moreo, Phys. Rep.
\textbf{344,} 1 (2001).

\bibitem{yarkony1996diabolical}D. R. Yarkony, Rev. Mod. Phys. \textbf{68,}
985 (1996).

\bibitem{Marcus} R. A. Marcus, Rev. Mod. Phys. 65, 599 (1993).

\bibitem{Tisdale11} W. A. Tisdale and X.-Y. Zhu, PNAS \textbf{108,}
965 (2011).\end{thebibliography}
\end{document}